\documentclass[twocolumn,letter]{jpsj2}
\usepackage{bm}

\title{%
Quantum Metamagnetic Transitions Induced by Changes in Fermi-Surface Topology $-$Applications to a Weak Itinerant-Electron Ferromagnet; ZrZn$_2$
}

\author{%
Youhei \textsc{Yamaji}\thanks{E-mail:yamaji@solis.t.u-tokyo.ac.jp}
,Takahiro \textsc{Misawa}, and Masatoshi \textsc{Imada}  
}

\inst{
 {\it 
Department of Applied Physics, University of Tokyo, 7-3-1 Hongo, Bunkyo-ku, Tokyo, 113-8656, Japan}
}

\recdate{March 7, 2007}

\abst{
We clarify
that metamagnetic transitions in three dimensions show unusual properties as
quantum phase transitions
if they are
accompanied by changes in Fermi-surface topology.
An unconventional universality deeply affected by the topological nature of Lifshitz-type transitions emerges around
the marginal quantum critical point (MQCP).
Here the MQCP is defined by the meeting point of the finite temperature critical line and 
a quantum critical line running on the zero temperature plane. The MQCP offers a marked contrast with the Ising universality and the gas-liquid-type criticality satisfied for conventional metamagnetic transitions.  At the MQCP, the inverse magnetic susceptibility $\chi^{-1}$ has diverging slope as a function of the magnetization $m$ (namely, $\left| d\chi^{-1}/dm \right|\rightarrow \infty$) in one side of the transition, which should not occur in any conventional quantum critical phenomena.
The exponent of the divergence can be estimated even at finite temperatures. We propose that such an unconventional universality indeed accounts for the metamagnetic transition in ${\rm ZrZn_{2}}$. }

\kword{
Lifshitz transition, Fermi-surface topology, metamagnetic transition, ${\rm ZrZn_{2}}$
}

\begin{document}
\maketitle
%
%
Itinerant ferromagnets such as ZrZn$_2$~\cite{Kimura04,Uhlarz04}, UGe$_2$~\cite{Huxley00} and nearly ferromagnetic metals such as Sr$_3$Ru$_2$O$_7$~\cite{Perry01
} show metamagnetic transitions. The magnetizations show jumps at magnetic fields separating the low-field phase with a smaller magnetic moment from the high-field phase with a higher moment.  The first-order transition characterized by the magnetization jump terminates at a finite-temperature critical point.  The universality around the critical point is regarded as the Ising type, which is equivalent to the gas-liquid critical points. The critical temperature can, however, be controlled to zero, for example, by pressure, which offers a field of quantum critical phenomena.  The metamagnetic transition and its quantum critical point (QCP) in itinerant electron systems have attracted interest because of its intriguing nature of fluctuations leading to possible non-Fermi-liquid behavior as well as superconductivity found in UGe$_2$ near the metamagnetic transition~\cite{Saxena00}.  

A simple scope of the QCP is offered by suppressions of critical temperatures of gas-liquid (or equivalently Ising) transitions, $T_{c}$, by quantum fluctuations~\cite{Sachdev99}.  The QCP of magnetic transitions in itinerant electron systems has been analyzed by the same framework of the suppressed symmetry breaking by the quantum fluctuations~\cite{Moriya85,Hertz76,Millis93}, which is expressed essentially by the $d+z$ dimensional Ising criticality with the spatial dimensionality $d$ being added by the dynamical exponent $z$ representing the quantum dynamics. The quantum criticality of the metamagnetic transition has been interpreted so far by the same conventional QCP of the Ising type~\cite{Millis02}.
\par
\begin{figure}[t]
\begin{center}
\includegraphics[width=7cm]{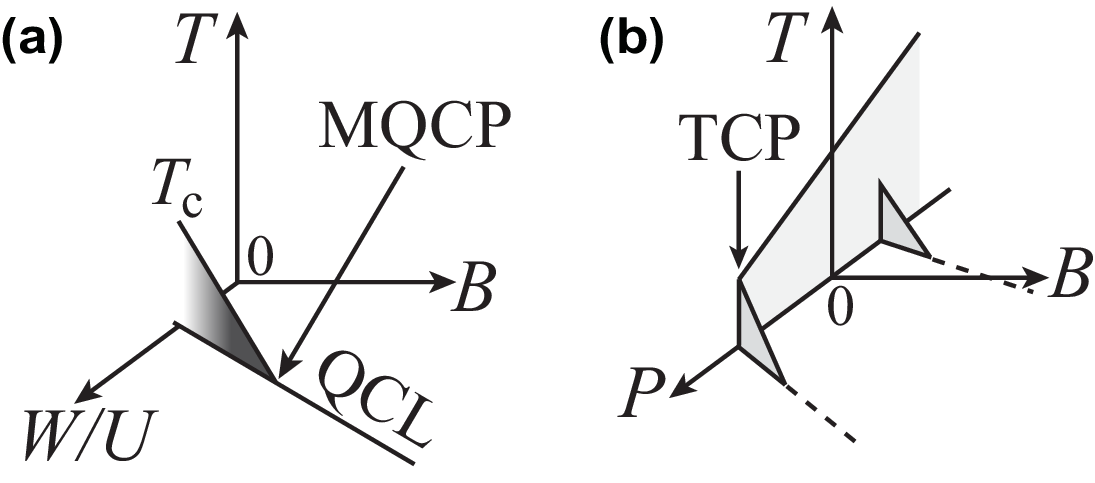}
\caption[]{(a) Schematic phase diagram around the marginal quantum critical point (MQCP), where a quantum critical line (QCL) and a finite-temperature critical line at $T_c$ meet. Here, $B$ controls quantum fluctuations
and corresponds to, for instance, magnetic field. Experimentally, pressure $P$ may control the ratio of itinerancy (bandwidth)$W$ to the interaction $U$. (b) Schematic phase diagram proposed for ${\rm ZrZn_{2}}$~\cite{Kimura04}. An extended range of applied pressure is illustrated including the negative side of $P$, where an additional metamagnetic transition has been speculated.  TCP represents the tricritical point, where the critical line at $B=0$ terminates and the transition becomes first order at $B=0$ below the tricritical temperature. This first-order line below TCP is the edge of the metamagnetic ``wing'', which is the main subject of this letter. The dashed lines were regarded just as a crossover in Ref. \citen{Kimura04}
based on the picture of the conventional QCP.} 
\label{diagram}
\end{center}
\end{figure}

Recently, however, for two-dimensional systems, a completely different type of quantum critical phenomena has been proposed for the metal-insulator transitions~\cite{Imada04,Misawa06} as well as for the Lifshitz transition~\cite{Yamaji06}, where the topological nature of the transition deeply modifies the above conventional picture.
In this case, the phase diagram is qualitatively different from that of the usual QCP, as is shown in Fig.\ref{diagram}(a).
Although the critical line at nonzero critical temperature $T_{c}$ is characterized by the conventional Ising universality,
there emerges a 
quantum critical line (QCL) of topological transitions starting from the QCP on the zero-temperature plane. This QCP is called {\it marginal quantum critical point} (MQCP) as the zero-temperature critical line terminates at this point. 

In this letter, we propose that the metamagnetic transition in three dimensional systems also bears an unconventional feature characterized neither by the Ising universality nor by any other universality of symmetry-breaking transitions, when the metamagnetic transition is involved and triggered by the topology change of the Fermi surface (ex. collapse of a neck of the Fermi surface or emergence/disappearance of a Fermi pocket, namely the Lifshitz transitions~\cite{Lifshitz60}).  This class of metamagnetic critical point is a manifestation of the MQCP. We propose that the metamagnetic transition of ZrZn$_2$ indeed belongs to this unconventional class.
The structure of the phase diagram proposed for ZrZn$_2$~\cite{Kimura04,Uhlarz04} shown in Fig.\ref{diagram}(b) is consistent with our interpretation shown in Fig.\ref{diagram}(a) for the part of the metamagnetic transition beyond the tricritical point (TCP). Furthermore, we propose that the dashed lines in  Fig.\ref{diagram}(b) are not crossover lines as in the interpretation in Refs.\citen{Kimura04}, and \citen{Uhlarz04} but represent real phase transition boundaries (QCL) of Lifshitz-type topological transitions.  We show that the magnetic field dependence of the inverse magnetic susceptibility $\chi^{-1}$ is a key quantity for the identification of the universality of the MQCP, where the slope diverges toward the MQCP. $\chi^{-1}$ observed for ZrZn$_2$ by Uhlarz et al.~\cite{Uhlarz04} indeed shows a remarkable enhancement. Even at finite temperatures, we show that the MQCP universality is easily observable beyond a crossover from the Ising one.

\par
We first discuss the itinerant electron metamagnetism by using 
a mean-field theory after Wohlfarth, Rhodes~\cite{Wohlfarth62}, and Shimizu~\cite{Shimizu82}(WRS). Through a mean-field decoupling of the Hubbard hamiltonian in the standard notation
\begin{align}
\hat{\mathcal{H}}=\sum_{
\bm{k}
\sigma}\left(\varepsilon (\mbox{\boldmath{$k$}})-\eta \frac{g\mu_{B}}{2}B \right)\hat{c}_{\bm{k}\sigma}^{\dagger}\hat{c}_{\bm{k}\sigma}^{\ }+\sum_{i}U\hat{n}_{i\uparrow}\hat{n}_{i\downarrow},
\end{align}
where $\eta$ is +1 (-1) for the spin $\sigma=\uparrow (\downarrow)$,
mean-field band dispersion for partially polarized quasiparticles, 
\begin{align}
E_{\bm{k}\sigma}=\varepsilon (\mbox{\boldmath{$k$}})-\eta\left(Um+g\mu_{B}B\right)/2,\label{Eq2}
\end{align}
are obtained. The magnetization, $m=n_{\uparrow}-n_{\downarrow}$ is selfconsistently determined afterwards. Here, the electron densities of each spin sub-bands, $n_{\uparrow},n_{\downarrow}$, are given as
\begin{align}
n_{\sigma}=\int_{-\infty}^{+\infty}\rho (\varepsilon) \left\{1+\exp\left( (\varepsilon-\mu_{\sigma})/T\right)\right\}^{-1}d\varepsilon,\label{Eq3}
\end{align}
where $\rho (\varepsilon)$ is the density of states (DOS) for the bare dispersion, $\varepsilon (\mbox{\boldmath{$k$}})$ and $\mu_{\sigma}=\mu + \eta ( Um+g \mu_{B} B ) / 2 $.
Then the magnetic-field dependence of the magnetization $m$ is self-consistently calculated from Eqs. (\ref{Eq2}) and (\ref{Eq3}) with $m=n_{\uparrow}-n_{\downarrow}$.
For the $canonical$ ensemble, which is justifiable under the charge neutrality, the total electron density, $n=n_{\uparrow}+n_{\downarrow}$ is fixed. In the following part of the present letter, the chemical potential $\mu$ is chosen to keep the electron density fixed.
\par
At zero temperature, the energy density
\begin{align}
f_{0}&=\sum_{\sigma}\int_{-\infty}^{\mu_{\sigma}}(\varepsilon -\mu_{\sigma})\rho (\varepsilon ) d \varepsilon +\frac{U}{4}(m^{2}+n^{2})\label{energy}
\end{align}
describes the nature of phase transitions.
In the present theory, the metamagnetic instability is signaled by the divergence of the uniform magnetic susceptibility, $\chi=\partial m/\partial B =2\rho_{{\rm eff}}(\mu)\left(1-U\rho_{{\rm eff}}(\mu)\right)^{-1}$ where $\rho_{{\rm eff}}$ is the effective DOS in the partially polarized state given as $\rho_{{\rm eff}}(\mu)=2\rho (\mu_{\uparrow})\rho (\mu_{\downarrow})/(\rho (\mu_{\uparrow})+\rho (\mu_{\downarrow}))$. The WRS theory is essentially based on the Ginzburg-Landau expansion of $f_{0}$ with respect to $m$, so-called the $\varphi^{6}$-theory.
However, we note that the Ginzburg-Landau expansion is not justified around the Lifshitz transition as we show below.
\par
%
\begin{figure}[t]
\begin{center}
\includegraphics[width=7cm]{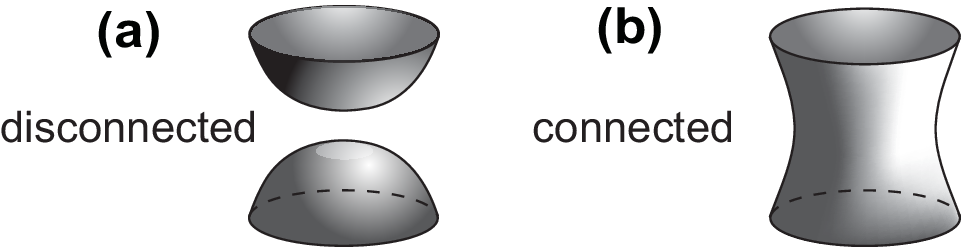}
\caption[]{(a) Fermi surface around a saddle point in the disconnected phase (see text). (b) Fermi surface in the connected phase.}
\label{fermi_surface}
\end{center}
\end{figure}
Now
we focus on the situation that necks of the Fermi surface can be
disconnected or connected
in 3D Brillouin zone
by changing
magnetic fields or applied pressure, as is expected in ${\rm ZrZn_{2}}$~\cite{Harima}.
In this case, a saddle point of the band dispersion, $\mbox{\boldmath{$k$}}_{L}$ is located near the Fermi surface.
Around the saddle point in three dimensions, we employ the axisymmetric expansion of the band dispersion as
\begin{align}
\varepsilon (\mbox{\boldmath{$k$}}) = \varepsilon (\mbox{\boldmath{$k$}}_{L})+\frac{\hbar^{2}}{2m_{\bot}^{\ast}}q_{\bot}^{2}+\frac{\hbar^{2}}{2m_{z}^{\ast}}q_{z}^{2}-\mu+\mathcal{O}(q^{4}),
\end{align}
where $\mbox{\boldmath{$q$}}=\mbox{\boldmath{$k$}}-\mbox{\boldmath{$k$}}_{L}$ and $m_{\bot}^{\ast}\cdot m_{z}^{\ast}<0$. The Fermi level at the Lifshitz transition is denoted as $\mu_{L}=\varepsilon (\mbox{\boldmath{$k$}}_{L})$ and the corresponding magnetic field as $B_{L}$. When the effective mass for the $xy$-plane, $m_{\bot}^{\ast}$ is positive (negative), we call the neck of the Fermi surface the electron (hole) neck. We also call the paramagnetic metallic phase with $(\mu-\mu_{L})m_{z}^{\ast}>0\ (<0)$, the disconnected (connected) phase (see Fig. \ref{fermi_surface}). At $\mu=\mu_{L}$, the density of states $\rho (\varepsilon)$ has a singularity originating from the changes in the Fermi-surface topology. For the connected-phase side $(\mu-\mu_{L})m_{z}^{\ast}<0$, the DOS $\rho (\varepsilon)$ may be expanded as
\begin{align}
\rho (\varepsilon)=\rho (\mu_{L})+\rho_{+}^{(1)}\zeta+\mathcal{O}(\zeta^{2}),\label{expand_0}
\end{align}
where $\zeta=\varepsilon-\mu_{L}$.
On the other hand, for the disconnected-phase side, $\rho (\varepsilon)$ may be expanded as
\begin{align}
\rho (\varepsilon)=\rho (\mu_{L})+\rho_{-}^{(1/2)}|\zeta|^{1/2}+\rho_{-}^{(1)}\zeta+
\mathcal{O}(\zeta^{3/2}),\label{expand_1}
\end{align}
which is analytic with respect to $|\zeta|^{1/2}$ except for the point $\zeta=0$.
The coefficient $\rho_{-}^{(1/2)}$ has a negative value.
\par

Substituting Eqs. (\ref{expand_0}) and (\ref{expand_1}) into Eq. (\ref{energy}), we expand
$f_{0}$ around the Lifshitz-transition point with respect to $\delta m \equiv -{\rm sign}(m_{z}^{\ast})(m-m_{L})$ and $\delta B=B-B_{L}$ as,
\begin{align}
f_{0}&\simeq {\rm const}+\frac{(1-U\rho_{{\rm eff}} (\mu_{L}))}{4\rho_{{\rm eff}}(\mu_{L})}\delta m^{2}
-\frac{g\mu_{B}}{2}\delta B
(m-m_{L})\nonumber\\
&+\left\{
\begin{array}{lc}
 c_{5/2}|\delta m|^{5/2}+\mathcal{O}(\delta m^{3})& (\delta m<0)\\
 c_{3}\delta m^{3}+\mathcal{O}(\delta m^{4})& (\delta m>0)
\end{array}
\right.,\label{Lifshitz}
\end{align}
where $c_{5/2}$
and $c_{3}$ are constants given by $\rho (\mu_{\uparrow}),\rho (\mu_{\downarrow}), \rho_{\pm}^{(1)}$ and $\rho_{+}^{(1/2)}$.
The region $\delta m<0\ (\delta m>0)$ corresponds to the disconnected (connected) phase. Here,
the Lifshitz transition occurs at $\delta m =0$.
The nonanalytic term in Eq. (\ref{Lifshitz}) for the disconnected side, $c_{5/2}|\delta m|^{5/2}$, originates from the integral
of the form $\int_{0}^{U\delta m} d \varepsilon \varepsilon  |\varepsilon  |^{1/2}$, which inevitably appears in
$f_{0}$ substituted with Eq. (\ref{expand_1}).
\par
Generally, the Lifshitz transitions may occur independently of possible metamagnetic transitions at zero temperature.
However, if $\rho_{{\rm eff}}$ has a maximum at the Lifshitz-transition point $\mu =\mu_{L}$ as a function of $\mu$, $\chi\propto (1-U\rho_{{\rm eff}}(\mu))^{-1}$ given below Eq.(\ref{energy}) diverges first at $\mu =\mu_{L}$ with the increase of $U$. Then a metamagnetic transition may occur simultaneously with the Lifshitz transition at $\mu =\mu_{L}$.
If $1/\rho_{{\rm eff}}(\mu_{L}) < U$ and $B\neq 0$ are satisfied,
this simultaneous transition occurs
as a first-order transition.
The first-order transition terminates at a nonzero critical temperature $T_{c}$, which has the conventional Ising universality.
The first-order transition terminates at $T=0$ for $U\rho_{{\rm eff}}(\mu_{L})\rightarrow 1$, which is nothing but the MQCP.
The Lifshitz-type QCL starts from the MQCP in the region $U\rho_{{\rm eff}}(\mu_{L})<1$.
The emergence of MQCP and first-order transitions is a consequence of the electron correlation effect,
while the Lifshitz transitions on the QCL remain ``transitions of the $2\frac{1}{2}$ order'' proposed for non-interacting systems by Lifshitz~\cite{Lifshitz60}.

\par

From the minimization of $f_{0}$ in Eq. (\ref{Lifshitz}), the singularity of the magnetization around the MQCP is obtained as
\begin{align}
\delta m \propto \left\{
\begin{array}{lc}
-
|B_{L}-B|^{1/\delta_{<}} & (B<B_{L}) \\
|B-B_{L}|^{1/\delta_{>}} & (B>B_{L})
\end{array}
\right. ,\label{s_delta}
\end{align}
where, for the electron-necks case $m_{z}^{\ast}<0$ (hole-necks case $m_{z}^{\ast}>0$),
the critical exponent $\delta$
is $\delta_{<}=3/2$ ($\delta_{<}=2$) for $B<B_{L}$ and $\delta_{>}=2$ ($\delta_{>}=3/2$) for $B>B_{L}$, which are clearly different from the mean-field Ising value, $\delta =3$.
In the case of the electron necks, the singularity of the inverse susceptibility $\chi^{-1}\equiv \partial B/\partial m$ is given as
\begin{align}
\chi^{-1}\propto
\left\{
\begin{array}{lc}
|B_{L}-B|^{1/3}\propto |\delta m|^{1/2}& (B<B_{L}) \\
|B-B_{L}|^{1/2}\propto |\delta m| & (B>B_{L})
\end{array}
\right. ,
\end{align}
whereas, in the case of the hole necks, $\chi^{-1}\propto |\delta m|$ for $B<B_{L}$ and $\chi^{-1}\propto |\delta m|^{1/2}$ for $B>B_{L}$.
At a conventional symmetry-breaking transition, $\chi^{-1}\propto |\delta m|^{\delta-1}$ vanishes with an exponent $\delta\geq 3$, 
resulting in a concave curve for the plot of $\chi^{-1}$ as a function of $m$. This applies to the present case with a nonzero $T_c$ 
as well.  In marked contrast, around the MQCP, $\chi^{-1}$ vanishes with a convex dependence with $\delta=3/2$ or at most linearly 
with $\delta=2$, which reflect the topological character of the transition. Along the QCL, on the other hand, $\chi^{-1}$ does not 
vanish.
%
\par
We now examine whether the MQCP is relevant to understand the experimental results for ${\rm ZrZn_{2}}$. 
${\rm ZrZn_{2}}$ is one of the canonical weak itinerant-electron ferromagnets with a small ordered moment 0.17$\mu_{B}$ per Zr atom and low Curie temperature $T_{c}=28.5$K, at ambient pressure.
Recent experimental results show that the magnetization suddenly
vanishes at $P=1.65$GPa and exhibits a first-order transition at $B=0$ below the temperature of a TCP~\cite{Uhlarz04}. In addition, two successive metamagnetic behaviors are detected for non-zero magnetic fields~\cite{Kimura04,Uhlarz04}, as shown in Fig.\ref{diagram}(b), where one of the transitions is connected to the TCP at $B=0$ and $P>0$.
\par
Several studies
have also been performed to elucidate the Fermi-surface topology of ${\rm ZrZn_{2}}$.
Singh and Mazin~\cite{Singh02} proposed the neck-free paramagnetic state, whereas electron-positron annihilation radiation measurements reveal the presence of the necks at ambient pressure~\cite{Major04}. 
A recent accurate {\it ab initio} electronic structure calculation by Harima shows that the necks around the $L$ point in the band closest to the Fermi surface change their topology depending on the location of the Fermi level~\cite{Harima}. These results together suggest that ${\rm ZrZn_{2}}$ is in a subtle proximity of the Lifshitz transition.
\begin{figure}[t]
\begin{center}
\includegraphics[width=8cm]{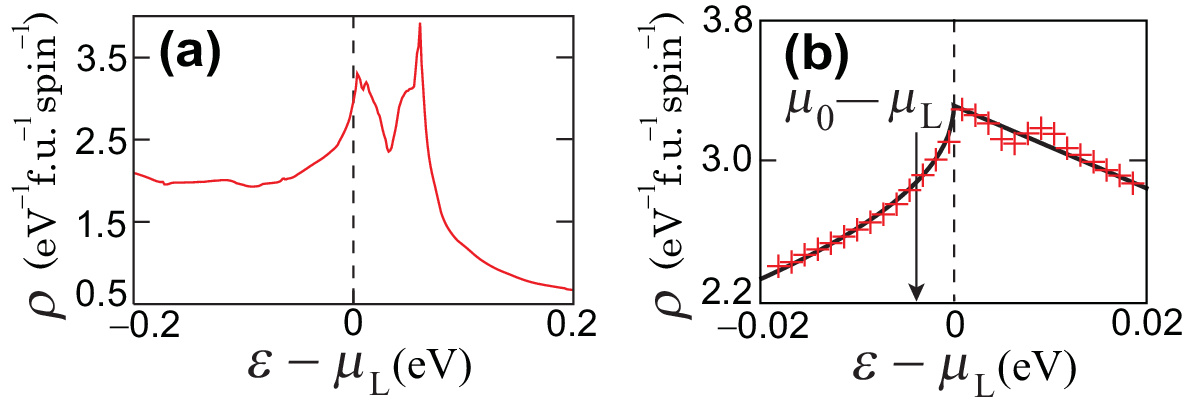}
\caption[]{
 (a) DOS $\rho (\varepsilon)$ of ${\rm ZrZn_{2}}$ calculated by using a full potential LAPW method based on LDA with lattice parameters in a paramagnetic phase, at ambient pressure~\cite{Harima}. (b) An enlargement of (a) around $\varepsilon =\mu_{L}$. The solid curve represents DOS fitted to the calculated result represented by the (red) cross, and is used in the present letter (see text). The Fermi level in the paramagnetic phase, $\mu_{0}$, is located below $\mu_{L}$.
}
\label{DOSv1}
\end{center}
\end{figure}
\par
Here we assume that an applied pressure just renormalizes the overall energy scale such as $U/W$ and leave the band structure calculated by LDA\cite{Harima} unchanged.
Then
saddle points associated with electron necks surrounding the $L$ point of the cubic Laves phase, around which quadratic band dispersions such as Eq. (5) are expected, are located above
the paramagnetic Fermi level as is shown in Fig. \ref{DOSv1}.
In this case, $\rho_{{\rm eff}}$ for ${\rm ZrZn_{2}}$ indeed has a maximum at the Lifshitz-transition point $\varepsilon=\mu_{L}$ as a function of $B$.
Hereafter we focus on this Lifshitz-transition point
and perform mean-field calculations by using the DOS estimated from fitting to the LDA results~\cite{Harima}, which is illustrated in Fig. \ref{DOSv1} (b). The fitting of the DOS is done by assuming the expansions of the DOS given by Eq. (\ref{expand_0}) and (\ref{expand_1}).
Here we ignore the fine structures of the DOS for $\varepsilon > \mu_{L}$ corresponding to other Lifshitz transitions.
Since the qualitative features of the MQCP do not depend on the precise location of the paramagnetic Fermi level $\mu_{0}$, we choose
$\mu_{0}$ at $\mu_{L}-2.7{\rm [meV]}$
to be consistent with the critical value of the magnetization expected from the experiment discussed below. 
\par
\begin{figure}[t]
\begin{center}
\includegraphics[width=8.5cm]{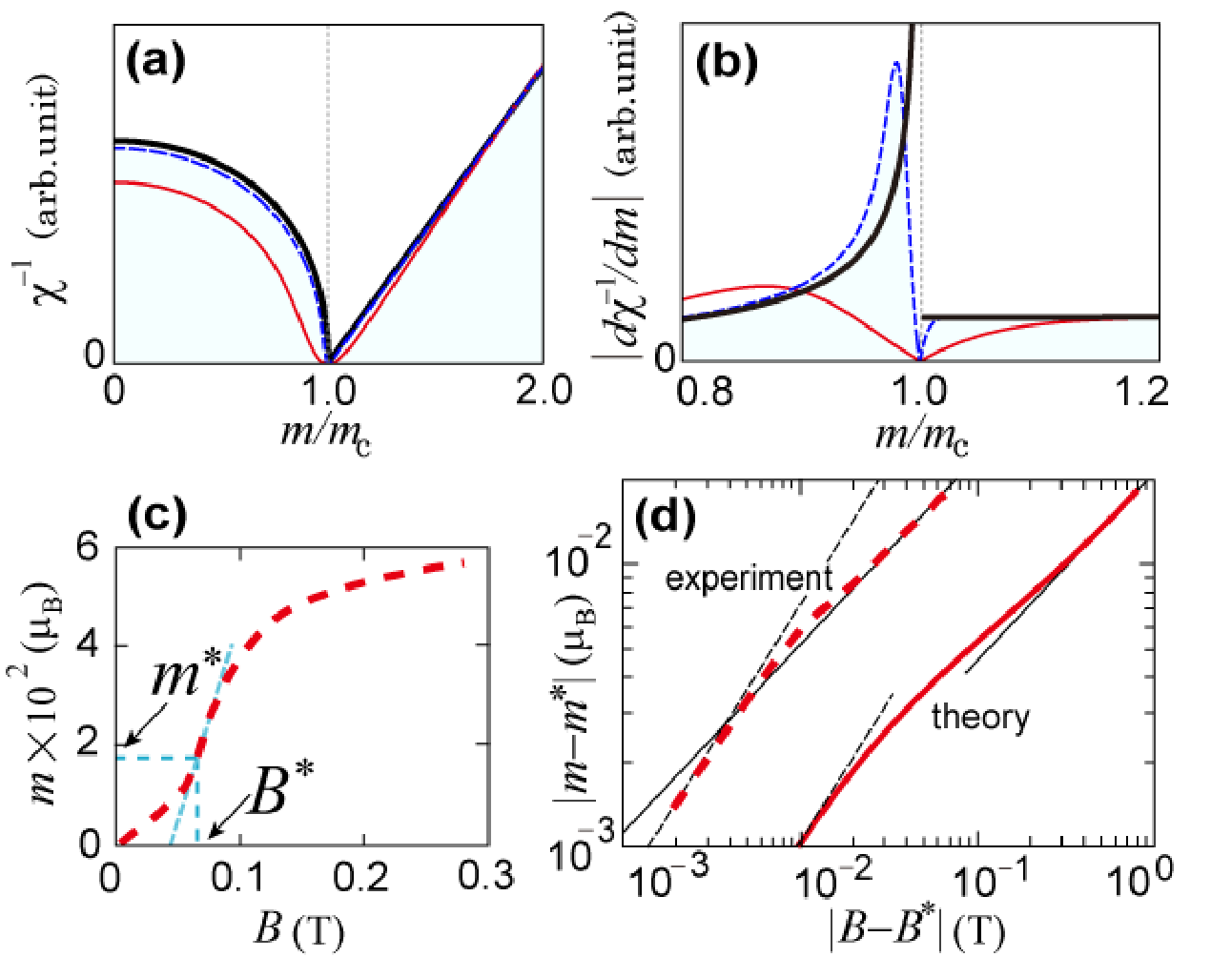}
\caption[]{(a) Inverse susceptibilities $\chi^{-1}$ as a function of $m$ calculated at fixed temperatures $T=T_{c}$. The thick solid curve represents $\chi^{-1}$ around the critical point at $T=T_{c}=0$ K, namely the MQCP. The thin dashed curve represents $\chi^{-1}$ at $T=T_{c}=0.15$ K and the thin solid curve represents $\chi^{-1}$ at $T=T_{c}=1.5$ K. (b) Absolute values of slope of inverse susceptibility $|d\chi^{-1}/dm|$ as a function of $m$.
The notations of the curves are the same as (a).
(c) Measured magnetization curve~\cite{Uhlarz04} at $P=2.1$ GPa and $T=3.0$ K. The arrows indicate $m^{\ast}$ and $B^{\ast}$. (d) Logarithmic plots of experimental and theoretical magnetization curves within the lower-field phase; the dashed and solid bold curves represent the experimental and theoretical magnetization, respectively. Straight solid lines represent $\delta=3/2$. The dashed straight lines represent the linear dependence; $m-m^{\ast}\propto B-B^{\ast}$.
}
\label{DOS}
\end{center}
\end{figure}
Although the MQCP emerges strictly at $T=0$,
the MQCP universality is observable even at finite temperatures beyond a crossover from the Ising one. The crossover is
 clearly seen in our calculated result for the $m$-dependence of $\chi^{-1}$ at a fixed $U/W$ and $T$. As is described above, the $m$-dependence of $\chi^{-1}$ is in sharp contrast with the Ising criticality: Within the lower-field phase, $\chi^{-1}$ shows a convex dependence on $\delta m$ and $|d\chi^{-1}/dm|$ diverges as $|m-m_{L}|^{-1/2}$ (see Fig.\ref{DOS}), whereas $\chi^{-1}$ for the Ising critical point should have
a concave dependence on $\delta m$ with vanishing $|d\chi^{-1}/dm|$. Even at finite temperatures, $|d\chi^{-1}/dm|$ is largely enhanced away form the critical point as is shown in Fig. \ref{DOS}. Here we choose the control parameter $U$ corresponding to $T_{c}$ lower than 1.5K.
As thermal fluctuations scaled by $T_{c}$ become suppressed, the enhancement of $|d\chi^{-1}/dm|$ becomes
significant 
for lower $T_{c}$ as is shown in Fig. \ref{DOS} (b).
\par
In Fig. \ref{DOS} (c) and (d), the magnetization curve of ${\rm ZrZn_{2}}$ at $P$=2.1 GPa and $T=$3.0 K reported in Ref.\citen{Kimura04} (see Fig. \ref{DOS} (c)) is compared with the mean-field result calculated at $T=3.0$ K for the parameter set corresponding to $T_{c}=1.5K$. Since the first-order transition appears to close just near this pressure, we have taken the crossover value $B^{\ast}$ which gives the maximum slope. Because 3.0 K
seems to be still higher than $T_{c}$ in the experimental data,
we expect that $|m-m^{\ast}|\propto |B-B^{\ast}|$ when $B$ is too close to $B^{\ast}$. The experimental data clearly show a crossover from $|m-m^{\ast}| \propto |B-B^{\ast}|$ to $|m-m^{\ast}| \propto |B-B^{\ast}|^{1/\delta}$ with $\delta >1$ at a very small value of $|B-B^{\ast}|\simeq 0.003$ T. The fitting of the experimentally observed scaling shows $\delta \simeq 1.5$ for $|B-B^{\ast}|>0.02$ T in remarkable agreement with the present prediction in Eq. (\ref{s_delta}) for MQCP shown as the solid curve in Fig. \ref{DOS} (d).
This suggests that the MQCP characterized by
$\delta =3/2$ must be located close to the point at $(m,B)=(m^{\ast},B^{\ast})$, where the fitting suggests $m^{\ast}$=0.018 $\mu_{{\rm B}}$ and $B^{\ast}$=0.066 T. 
It would be interesting to tune $P$ and $B$
to find the pinpoint of the MQCP.
\par
An experimental Stoner factor estimated by $S=\left.\chi\right|_{B=0}/2\rho (\mu_{0})$ becomes extremely large as $S\sim 10^{3}$, which appears to be beyond the applicability of the present mean-field treatment. This may originate from density fluctuations and may cause a parallel shift of the curve as is seen in Fig. \ref{DOS}(d), while the criticality and the crossover analyzed in this letter
should remain valid.
\par
The real transition expected all through along the QCL instead of the crossover may be detected by a jump of the thermopower~\cite{Vaks81}
in this quantum topological metamagnetism, although at low $T\neq 0$, the QCL becomes a sharp crossover.\par
We finally discuss the qualitative reliability of the present mean-field theory around the MQCP.
There are two possible limitations of the present theory; an instability of the homogeneous phase and influences of density fluctuations apart from the largely enhanced Stoner factor discussed above.
 Around the MQCP expected for ${\rm ZrZn_{2}}$, the linear instability of homogeneous phases signaled by the negative values of the uniform charge compressibility $\kappa \propto \left( 1-U\widetilde{\rho}_{{\rm eff}}\right)^{-1}$, where $\widetilde{\rho}_{{\rm eff}}=\sqrt{\rho (\mu_{\uparrow}) \rho (\mu_{\downarrow})}$~\cite{Yamaji06}, occurs in a tiny field range given by $|B-B_{L}|/B_{L}<5\times 10^{-4}$ for $B<B_{L}$ and $|B-B_{L}|/B_{L}< 10^{-2}$ for $B>B_{L}$, even if the costs of long-range Coulomb energy for phase separations are ignored.
Even when such a phase separation is expected, the enhancement of $|d\chi^{-1}/dm|$ remains observable outside the region of inhomogeneity, whereas the Ising criticality could be masked by the phase separation around the critical point located at low temperatures. The
density fluctuations do not affect the validity of the mean-field exponents for $B>B_{L}$, where the expansion Eq.(\ref{Lifshitz}) characterized by a $\varphi^{3}$-theory is on the upper critical dimension $d_{c}=d+z=6$. On the other hand, the upper critical dimension of the expansion Eq.(\ref{Lifshitz}) for $B<B_{L}$ may be higher than 6. Therefore, in the close proximity of the MQCP, $\delta >3/2$ could be obtained, although it is not seen in the experiment.
\par
In summary, the metamagnetic transition in three
dimensions accompanied by changes in the Fermi-surface
topology is shown to belong to an unusual universality class.
The topological and symmetry-breaking characters meet at the MQCP,
where the inverse susceptibility $\chi^{-1}$ has a divergent slope
as a function of the magnetization $m$, in contrast with the
Ising criticality. The experimental results for the metamagnetic transition of ${\rm ZrZn_{2}}$ are consistent with this prediction, 
implying the relevance of the unconventional universality.
\par
We would like to thank H. Harima for providing us with his detailed calculated data on the band structure of ZrZn$_2$. This work is supported by Grant-in-Aids for Scientific Research on Priority Areas under the grant numbers 17071003 and 16076212 from MEXT, Japan.

\end{document}